\documentstyle[prl,aps,twocolumn]{revtex}
\begin{document}
\draft
\title{Charge Transfer Induced Persistent Current
and Capacitance Oscillations}
\author{M.\ B\"uttiker$^{\ast}$ and C.\ A.\ Stafford$^{\dagger}$}
\address{D\'epartement de physique th\'eorique,
Universit\'e de Gen\`eve,
CH-1211 Gen\`eve 4, Switzerland}
\address{\rm (4 September 1995)}
\address{\mbox{ }}
\address{\parbox{14cm}{\rm \mbox{ }\mbox{ }\mbox{ }
The transfer of charge between different regions of a
phase-coherent mesoscopic sample is investigated.
Charge transfer from a side branch quantum dot into a ring
changes the persistent current through a sequence of plateaus
of diamagnetic and paramagnetic states.
In contrast, a quantum dot embedded in a ring
exhibits sharp resonances in the persistent current, whose sign
is independent of the number of electrons
in the dot if the total number of electrons in the system is even.
It is shown that such a mesoscopic system can be polarized appreciably
not only by the application of an external voltage, but also via an
Aharonov-Bohm flux.
}}
\address{\mbox{ }}
\address{\parbox{14cm}{\rm PACS numbers:  73.20.Dx, 73.40.Gk, 71.27.+a}}

\maketitle

\makeatletter
\global\@specialpagefalse
\def\@oddhead{\underline{REV\TeX\mbox{ }3.0\hspace{11.7cm}
UGVA-DPT 1995 / 09-902}}
\let\@evenhead\@oddhead
\makeatother

\narrowtext

The transfer of a single electronic charge from one region
of a mesoscopic conductor into another region of the
conductor can dramatically alter the mesoscopic properties
of the conductor. In this work, we take the persistent
current of a ring\cite{MBYIL,LEVYD,BOUCH}
as a phase sensitive probe of the equilibrium state of
the conductor and investigate its properties under
charge transfer. In Fig.\ 1,
two samples are shown in which a ring-like structure
is penetrated by an Aharonov-Bohm flux $\Phi$ and is connected
to a quantum dot. If the sample is brought into an
external capacitive circuit it can be polarized;
charge transfer from one portion of the sample into the
quantum dot can be induced. The charge transfer changes
the potential landscape, and with it changes the phase sensitive
properties of the mesoscopic sample.
Both the electrochemical
capacitance $C_{\mu} = e d\langle Q \rangle/d\mu$ and the
flux-induced capacitance $C_{\Phi} = e d\langle Q\rangle/d\Phi$ are
periodic functions of the AB-flux\cite{BU94}.
For the samples of Fig.\ 1, we find indeed very striking flux sensitive
features in these capacitance coefficients. Measurement of such capacitance
coefficients provides an important alternative to
the difficult magnetization measurements\cite{LEVYD}
used to characterize the
ground state of mesoscopic samples.

A purely one-dimensional ring exhibits a persistent
current which is either diamagnetic or paramagnetic
depending on the number of particles and their
distribution over the flux sensitive states\cite{LOSS}.
The persistent current is always an odd function of flux
$I(\Phi) = - I (-\Phi)$. But the slope of the persistent
current $dI(\Phi)/d\Phi$ for a small flux can be either negative
(diamagnetic) or positive (paramagnetic).
To be brief, we say that a diamagnetic ground state
has a positive parity and a paramagnetic state has
a negative parity.  If we consider the contribution
to the persistent current of each spin class separately,
then the addition of a single electron changes the parity of its
spin class\cite{LOSS}.
For the sample in Fig.\ 1a, in which the dot acts as a fully coherent reservoir
of carriers, charge transfer thus induces sharp transitions between plateaus
of diamagnetic and paramagnetic states.

For the ring of Fig.\ 1b, the persistent current is suppressed by charging
effects unless the conditions for resonant charge transfer are met,
an effect analogous to the {\em Coulomb blockade} observed in
the conductance through a
quantum dot coupled to macroscopic leads \cite{kast1,block1,mwl}.
The charge transfer discussed here should, however, be distinguished from
the standard discussions of the Coulomb blockade \cite{block1},
which treat charge transfer incoherently.
Here we deal with coherent many-body states which are extended
over {\em multiple} regions\cite{sandi,klimeck,2dots}.
The surprising effect which we find for the ring of Fig.\ 1b is that
the sign of the persistent current contributed by each spin class is
{\em independent} of the number of electrons in the dot.
The parity of each spin class is conserved under charge transfer, and
is determined only by the total number of electrons
in the sample, regardless of whether these electronic
\begin{figure}[htb]
\vbox to 5.0cm {\vss\hbox to 8cm
  {\hss\
    {\includegraphics{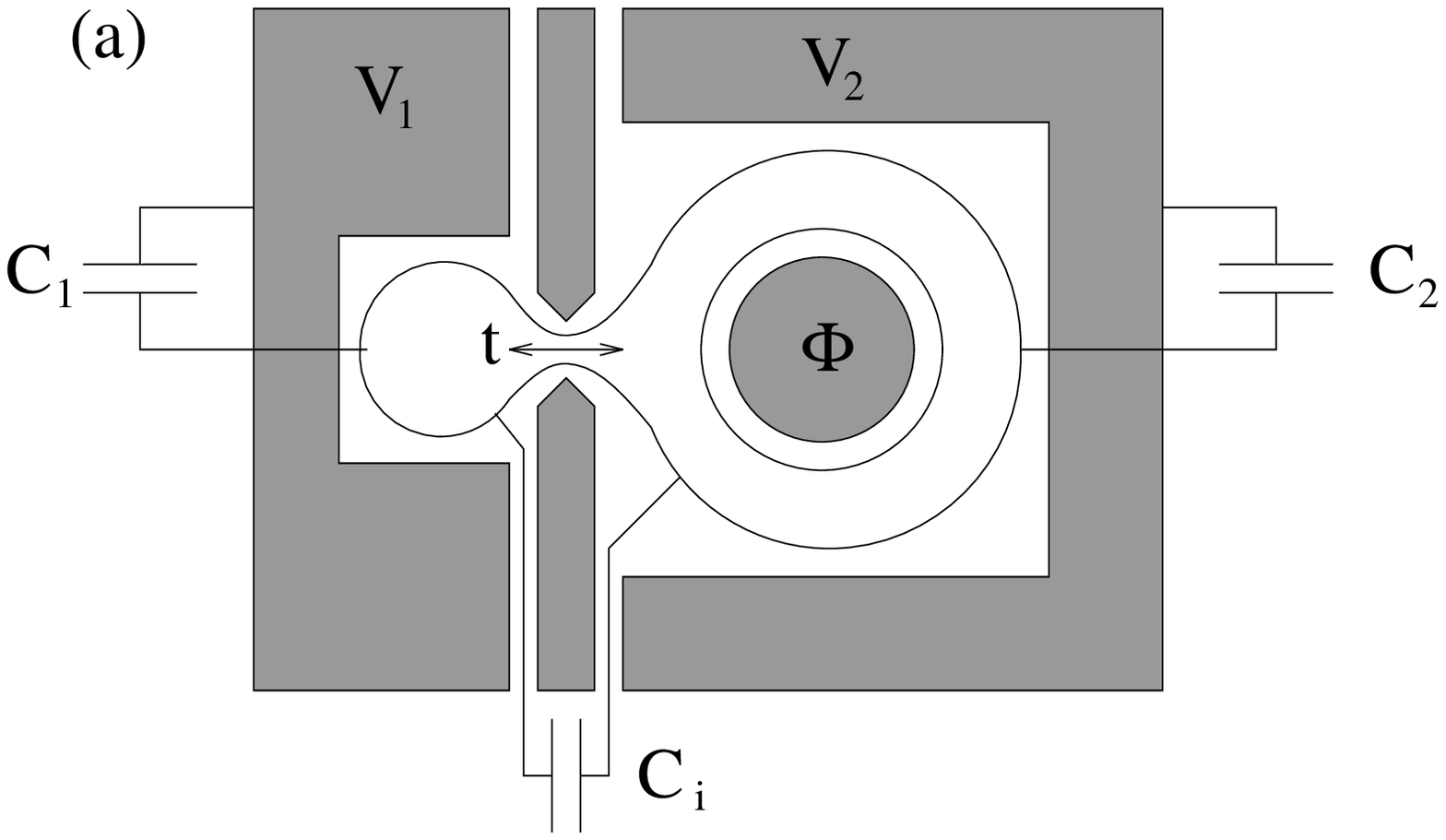}}
   \hss}
}
\vbox to 4.5cm {\vss\hbox to 8cm
  {\hss\
    {\includegraphics{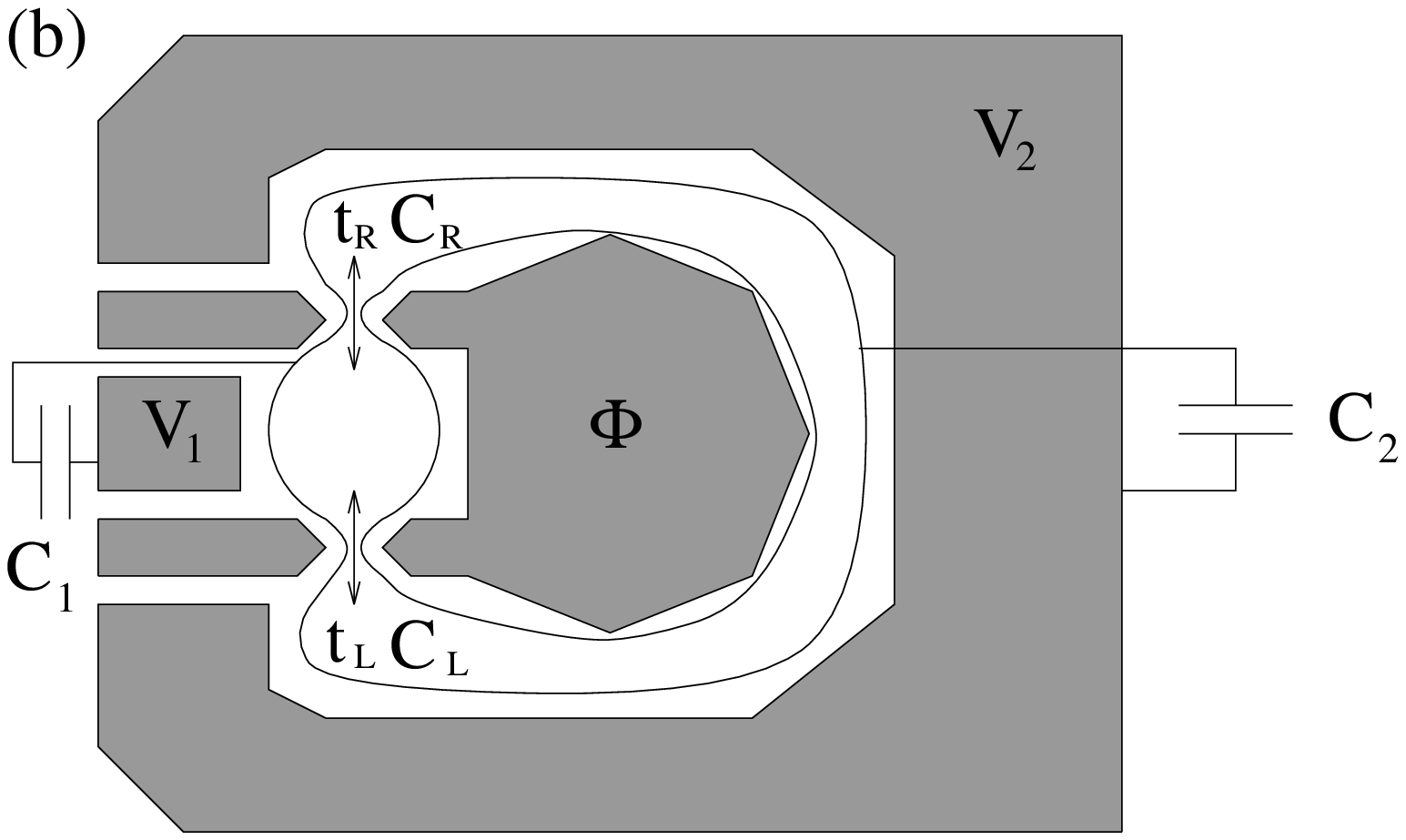}}
   \hss}
}
\caption{(a) Ring with Aharonov-Bohm flux coupled to a side branch quantum dot.
(b) Quantum dot with leads closed into a loop.}
\label{fig1}
\end{figure}
\noindent
states are localized or whether the states are extended and
contribute to the persistent current. In contrast, the sample
in Fig.\ 1a changes its parity with each electron that
is transferred from the dot into the ring.

The geometry of Fig. 1a has been the
subject of Refs. \cite{BU94,JAYA}.  A recent experiment\cite{YACO}
and theory\cite{Levy} investigated the AB effect of
a quantum dot embedded in a loop and connected
to two leads.  Here we treat explicitly the
{\em capacitively-coupled} closed structures and analyze
the charge response.

The system of Fig.\ 1 is modeled in terms of a one-dimensional ring
coupled capacitively and via tunneling to a quantum dot.
The electron-electron interactions in the system are treated using a
capacitive charging model, as indicated in Fig.\ 1:
The system is coupled
to two external metallic gates at voltages $V_1$ and $V_2$ with
capacitance coefficients
$C_1$ and $C_2$.  In addition, the quantum dot couples to the ring with
capacitance $C_i$ ($\mbox{}=C_R+C_L$ for the case shown in Fig.\ 1b).
With the combined capacitances
$C^{-1}_{e} = C^{-1}_{1} + C^{-1}_{2}$ and $C = C_{e} + C_{i}$,
we can express the electrostatic Hamiltonian (which includes the work done
by the voltage sources $V_1$ and $V_2$) in terms of the charge operator
for the dot $Q=\sum_{n\sigma}d^{\dagger}_{n\sigma}d_{n\sigma}$,
a polarization charge $Q_{0} =  C_{e} V$, and the
externally applied voltage $V = V_{2}- V_{1}$,
\begin{eqnarray}
H_{C} = (1/2C) (Q-Q_{0})^{2} - (C_{e}/2) V^{2}.
\label{hc}
\end{eqnarray}
The total Hamiltonian for the system is
$H = H_0 + H_T + H_C$, where
$H_0 = \sum_{k\sigma} \epsilon_{ak} c^{\dagger}_{k\sigma}c_{k\sigma}
+ \sum_{n\sigma} \epsilon_{dn} d^{\dagger}_{n\sigma}d_{n\sigma}$
describes the single-particle eigenstates in the ring and the dot, and
the tunneling Hamiltonian is
$H_T =
 \sum_{kn\sigma} \left(t_{kn} d^{\dagger}_{n\sigma} c_{k\sigma}
+ \mbox{H.c.}\right)$.
For the system of Fig.\ 1a, the AB-flux modulates the single-particle
energy levels $\epsilon_{ak}$ in the ring, while for the system of Fig.\ 1b,
the tunneling matrix elements $t_{kn}$ connecting the dot to the ring are
flux dependent.  $H_C$ favors integer
charge states of the quantum dot \cite{kast1,block1,mwl}, whereas
$H_T$ promotes hybridization of the localized states on the dot with the
extended states of the ring.  Our Hamiltonian is similar to
the Anderson model \cite{pwa} for a magnetic impurity (or quantum dot
\cite{kondot}) coupled to a Fermi sea of conduction electrons, but here the
reservoir of conduction electrons is itself a mesoscopic system with a
finite level spacing and bandwidth.  In order to account for the tendency
toward charge quantization in the system,
$H_C$ must be treated nonperturbatively.  We therefore
employ two complementary approaches:
In the weak-tunneling limit, where hybridization occurs only
between a single state in the ring and in the dot, $H$ can be reduced to
a $2\times 2$ matrix ($3\times 3$ including spin), allowing for an explicit
solution.  This simple analytical solution
correctly describes the interesting parity effects in the system.
The ground state is also found exactly for arbitrary coupling using a numerical
Lanczos technique.

Figs.\ 2a and b show numerical results for the
persistent current $I=-cdE_0/d\Phi$ and the electrochemical capacitance
$C_{\mu}=-d^2E_0/dV^2$ of the systems of
Figs.\ 1a and b, respectively,
as a function of the polarization charge $Q_0$.  Here
$E_0$ was evaluated computationally, with the single-particle
energy levels $\epsilon_{ak}$ and $\epsilon_{dn}$ in the ring and dot
and the tunneling matrix elements $t_{kn}$
modeled using a one-dimensional
tight-binding model in which the dot was represented by 2 sites, and the
ring by 4 sites.  The matrix element $w$ of the kinetic energy operator
between nearest-neighbor sites within the ring and the dot was taken
to be unity, and the point contacts were modeled as weak links.
For the case of 3 up-spin and 3 down-spin
electrons, the persistent current of the quantum dot within
the loop is diamagnetic, while the loop with a side branch quantum dot
exhibits a sequence of plateaus of diamagnetic and paramagnetic states.
The four peaks in $C_{\mu}$ in Figs.\ 2a,b, separated by $\Delta Q_0 \sim e$,
correspond to the successive transfer of
electrons from the ring to the dot (for decreasing $Q_0$), filling the four
available single-particle states in the dot.

In order to understand the character of the charge
transfer induced oscillations in $I$ and $C_{\mu}$, it is useful
to consider the limit
$t_{kn} \ll \Delta \epsilon,\; e^2/C$, where the dot and the ring are only
weakly coupled.  Then, in the vicinity of the
charge transfer resonance $N \rightarrow N+1$, where $N$
\begin{figure}[htb]
\vbox to 8.7cm {\vss\hbox to 8cm
  {\hss\
    {\includegraphics{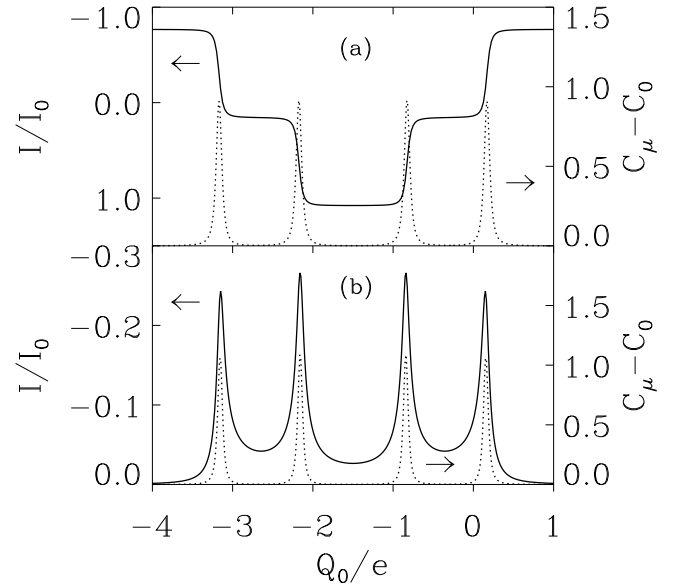}}
   \hss}
}
\caption{Persistent current and differential capacitance as a function of the
polarization charge $Q_0 =C_e V$ for (a) the sample of Fig. 1a with $t=0.5$
and (b) the sample of Fig. 1b with $t_R=0.2$ and $t_L=0.3$.
Each sample contains 6 electrons, with $\Phi/\Phi_0= 1/4$, and
$e^2/C=10$. Energy is expressed in units of $w$, $4w$ being
the bandwidth in the ring.
The persistent current (solid curve) is expressed in units of $I_0=
\max e v_F/L$, where $L$ is the circumference of the ring, and the capacitance
(dotted curve) is expressed in units of $(C_e/C)^2 e^2/w$.
Each peak in $C_{\mu}$ corresponds to the transfer of one electron
from the dot to the ring.}
\label{fig2}
\end{figure}
\noindent
is the number of electrons in the dot,
one need only consider the hybridization of the
highest occupied level $|aM\rangle$ in the
ring with the lowest unoccupied level $|d(N+1)\rangle$ in the dot.
Neglecting spin (the effects of which will be considered further below),
the Hamiltonian then reduces to a $2\times2$ matrix,
\begin{equation}
H_{\mbox{h}}=
\left(\begin{array}{cc} \epsilon_{aM} + \frac{(eN+Q_0)^2}{2C} & t\\
t^{\ast} & \epsilon_{d(N+1)} + \frac{[e(N+1)+Q_0]^2}{2C}
\end{array}\right)
\label{2by2}
\end{equation}
plus an additive constant
\begin{equation}
E_1 = \sum_{k=1}^{M-1} \epsilon_{ak} + \sum_{n=1}^{N} \epsilon_{dn} -C_e V^2/2,
\label{constant}
\end{equation}
where $M+N$ is the total number of (spinless) electrons in the system.
For the system of Fig.\ 1b, the matrix element $t$ depends on the total
number of nodes $M+N-1$ in the wave functions $|aM\rangle$ and
$|d(N+1)\rangle$:
its modulus squared is given by
\begin{equation}
|t_{\pm}|^2 = t_R^2 + t_L^2 \pm 2t_R t_L \cos(2\pi\Phi/\Phi_0),
\label{tpm}
\end{equation}
where the $+$ sign holds for $M+N-1$ even, and the $-$ sign holds for
$M+N-1$ odd.  Here $\Phi_0= h c/e$
is the single-charge flux quantum and $t_{R/L}$ are energies proportional
to the transmission amplitudes through the two point contacts.
The hybridization of the localized state of the dot with the extended state
of the ring is a maximum when the polarization charge takes the value
\begin{equation}
Q_{\ast} = -e(N+1/2) + (C/e)[\epsilon_{aM} - \epsilon_{d(N+1)}].
\label{qhyb}
\end{equation}
Note that this is precisely the polarization charge which would be needed
to transfer an electron in the classical approach
to the Coulomb blockade. For this polarization charge, in the classical
case, the energy has the form of a cusp.
In the quantum mechanical case, the ground state
energy is a smooth function of the
polarization charge,
\FL
\begin{eqnarray}
E_0 = E_1 & + & \frac{\epsilon_{aM} + \epsilon_{d(N+1)}}{2}
+ \frac{e^2}{8C} + \frac{[e(N+1/2)+Q_0]^2}{2C}
\nonumber \\
& - & \frac{1}{2}
\left(\left[\frac{e}{C}(Q_0-Q_{\ast})\right]^2 + 4 |t_{\pm}|^2\right)^{1/2}
\!\!\!\!\!\!.
\label{e0}
\end{eqnarray}
Due to quantum mechanical tunneling, the energy barrier is
lower. Note that after transfer of an electron to the dot the next
hybridization will take place between the state $|a(M-1)\rangle$ of the ring
and the state $|d(N+2)\rangle$ of the dot.
The total number of nodes $(M-1)+(N+2)-2 = M+N-1$,
which determines the parity of the system, is left invariant.
Let us next explore a few consequences of this simple result.

Differentiating Eq.\ (\ref{e0}), one obtains the persistent current
for the sample of Fig. 1b,
\begin{eqnarray}
I (\Phi) = \mp \frac{e}{\hbar}\frac{4 \pi t_{R}t_{L}\sin(2\pi\Phi/\Phi_0)}{
\left([e(Q_0-Q_{\ast})/C]^{2}+ 4|t_{\pm}|^{2}\right)^{1/2}}.
\label{current}
\end{eqnarray}
The persistent current is a sharply peaked function of the
polarization charge, obtaining a maximum value of $I_{\max}=c\,\partial
|t_{\pm}|/\partial \Phi$ at $Q_0 = Q_{\ast}$, and being of order
$(e/\hbar)[t_R t_L/(e^2/C)]$ far from resonance.
The parity of $I(\Phi)$ is determined
by the matrix element $t_{\pm}$, and is independent of the polarization
charge $Q_0$.  This result is a consequence of the Friedel sum rule\cite{Levy},
which
links the phase acquired by an electron traversing the system to the {\em
total}
charge in the system, which is invariant under polarization.
Consequently,
the parity effects on the persistent current described here for the case
of a strictly one-dimensional ring are expected to be quite general.
Eq.\ (\ref{current})
indicates that the peaks in the persistent current exhibit long
non-Lorentzian tails away from resonance due to charge fluctuations on
the quantum dot, as is evident in Fig.\ 2b.

The charge on metallic gate $1$ is determined by $Q_{e} = - dE_0/dV$.
The electrochemical capacitance between gates 1 and 2 is thus
$C_{\mu} = - d^{2}E_0/dV^{2}$. From Eq. (\ref{e0}), we find
\begin{eqnarray}
C_{\mu} - C_{0}= \frac{2 e^{2} (C^{2}_{e}/C^{2}) |t_{\pm}|^2}{
\left([e(Q_0-Q_{\ast})/C]^{2}+ 4|t_{\pm}|^{2}\right)^{3/2}},
\label{cap}
\end{eqnarray}
where $C^{-1}_{0} = C^{-1}_{e}+C^{-1}_{i}$ is the classical series capacitance.
The total change of the charge on gate 1 integrated over such a charge
transfer resonance (excluding the contribution from $C_0$)
is $|\Delta Q_e| = e(C_e/C)$, corresponding to the
transfer of one electron between ring and dot.
The quantum corrections to $C_{\mu}$
reach a maximum of $(eC_e/2C)^2/|t_{\pm}|$
at $Q_0=Q_{\ast}$, and are of order $C[|t_{\pm}|/(e^2/C_e)]^2$ far from
resonance, decreasing faster than a Lorentzian.
The coherent backscattering in such a phase-coherent system thus leads to a
suppression of charge transfer away from resonance vis \`a vis
a system with incoherent
charge transfer, such as that studied by Ashoori {\it et al.} \cite{ashoori}
or Lafarge {\it et al.} \cite{LAFAR},
which would be expected to exhibit Lorentzian peaks at zero temperature.
That is to say, coherence
suppresses charge fluctuations of the type $\delta Q = \langle Q - Ne\rangle$,
which contribute to the capacitive response of the system, while
enhancing charge fluctuations of the type
$\delta Q = (\langle Q^2 \rangle -\langle Q \rangle^2)^{1/2}$, which govern
the persistent current.
The parity of $t_{\pm}$ determines the phase of the AB-effect on $C_{\mu}$,
which exhibits a phase-shift of $\pi$ on resonance.

Let us next briefly describe how the above results change for the loop
with the side branch quantum dot. If the loop and the side dot are
disconnected,
the ring supports flux dependent states with energies
$\epsilon_{ak}(\Phi)$ whereas
the dot supports flux independent states with energies $\epsilon_{dn}$.
Thus, for this system, we have a persistent current $I(\Phi)$ even in the
absence of coupling to the dot. To take the Coulomb interaction
into account in the presence of a weak coupling to the dot,
we again need only consider
the hybridization of the topmost electron in the ring with the lowest
empty state in the dot. For the energy of the topmost electron, this leads
to an eigenvalue problem of the same form
as Eq.\ (\ref{2by2}),
but now with coupling matrix elements $t$ which are independent of
flux. The total energy is of the same form as Eq.\ (\ref{e0}), except that
the flux dependence is now determined by the states of the uncoupled ring.

The sensitivity of the persistent current
to changes in the gate voltage can be characterized
by the flux-induced capacitance \cite{BU94} $C_{\Phi}$.
This capacitance is measured in response to an oscillating AB-flux
$d\Phi (\omega) \exp(-i\omega t)$ superimposed on the static AB-flux,
and is given by $C_{\Phi} = - d^2E/d\Phi dV = -(1/c)dI(\Phi)/dV$.
The flux-induced capacitance is, like the persistent current, an
odd function of flux.  It has a particularly interesting behavior for the
system of Fig.\ 1a, for which case it takes the form
\begin{equation}
C_{\Phi} = \frac{4t^2 e (C_e/C) d\epsilon_{aM}(\Phi)/d\Phi}{
(\{e[Q_0 - Q_{\ast}(\Phi)]/C\}^2 + 4t^2)^{3/2}}
\label{cphi}
\end{equation}
near resonance.  Because $Q_{\ast}$ is now a function of
the AB-flux $\Phi$, one can pass through the charge transfer resonance
by varying $\Phi$.
Integrating Eq.\ (\ref{cphi}) with respect to $\Phi$, one
finds $|\Delta Q_e| = e(C_e/C)$
for the case where the bandwidth in the ring is large compared to $t$,
corresponding to the transfer of one electron between ring and dot.

So far we have neglected spin. For the sample
of Fig.\ 1b, the discussion given above still applies in the vicinity
of a single resonance for the case
when there are an unequal number of up-spin and down-spin
electrons in the system.  However,
the parity of the persistent current on resonance is then determined by
the spin of the electron being transferred.  If the up-spin and down-spin
systems have different parity, this leads to resonances of {\em alternating}
sign in the persistent current.
For equal numbers of up-spin and down-spin electrons, the ground state
forms a Kondo singlet.  In the weak-coupling limit, the Hamiltonian reduces
to a tridiagonal
$3\times 3$ matrix similar to Eq.\ (\ref{2by2}), where the diagonal
terms give the energies of the three possible charge states in the absence
of tunneling, and the terms nearest the diagonal are $\sqrt{2}\, t_{\pm}$ and
$\sqrt{2}\, t_{\pm}^{\ast}$.
This leads to an enhancement of the persistent current on resonance by
a factor of $\sqrt{2}$ compared to Eq.\ (\ref{current}), and an enhancement
by a factor of 2 midway between the two resonances.  In such a system, the
parity of the persistent current is again invariant under charge transfer,
as illustrated in Fig.\ 2b.

The transfer of a single electronic charge from one region
of a mesoscopic conductor into another region of the
conductor can dramatically alter the mesoscopic properties
of the conductor. In this work we have taken the persistent
current as an example. We have emphasized that the measurement
of capacitance coefficients $C_{\mu}$ and $C_{\Phi}$ provides
an interesting possibility to characterize the ground state of such
closed systems.
The charge transfer in quantum-coherent mesoscopic
conductors or large molecules thus provides a very interesting future
avenue of research.

This work was supported by the Swiss National Science Foundation.

\end{document}